\documentclass[aps,prd,twocolumn,superscriptaddress,amsmath,amssymb,amsthm,nofootinbib, preprintnumbers]{revtex4}
\usepackage{url}
\usepackage[colorlinks=true,
    linkcolor=blue,
    citecolor=blue,
    urlcolor=blue]{hyperref}
 \usepackage{amsmath}
\usepackage{graphicx}
\usepackage{epstopdf}
\usepackage{float}
\usepackage{hyperref}
\usepackage{color}
\usepackage[T1]{fontenc}
\usepackage[utf8]{inputenc}
\usepackage[toc,page]{appendix}
\usepackage[usenames,dvipsnames]{xcolor}
\usepackage[normalem]{ulem}

\newcommand{\be}{\begin{equation}}
\newcommand{\ee}{\end{equation}}
\newcommand{\ba}{\begin{eqnarray}}
\newcommand{\ea}{\end{eqnarray}}

\newcommand{\beq}{\begin{equation}}
\newcommand{\eeq}{\end{equation}}
\newcommand{\beqa}{\begin{eqnarray}}
\newcommand{\eeqa}{\end{eqnarray}}

\begin{document}


\title{Rotating spacetimes with a free scalar field in four and five dimensions}

\author{Jos\'e Barrientos}
\email{jbarrientos@academicos.uta.cl}
\affiliation{Sede Esmeralda, Universidad de Tarapac{\'a}, Avenida Luis Emilio Recabarren 2477, Iquique, Chile}
\affiliation{Institute of Mathematics of the Czech Academy of Sciences, {\v Z}itn{\'a} 25, 115 67 Praha 1, Czech Republic}
\preprint{CERN-TH-2025-011}

\author{Christos Charmousis}
\email{christos.charmousis@ijclab.in2p3.fr}
\affiliation{Université Paris-Saclay, CNRS/IN2P3, IJCLab, 91405 Orsay, France}
\affiliation{Theoretical Physics Department, CERN, 1211 Geneva 23, 
Switzerland}

\author{Adolfo Cisterna}
\email{adolfo.cisterna@mff.cuni.cz}
\affiliation{Sede Esmeralda, Universidad de Tarapac{\'a}, Avenida Luis Emilio Recabarren 2477, Iquique, Chile}
\affiliation{Institute of Theoretical Physics, Faculty of Mathematics and Physics,
Charles University, V Hole{\v s}ovi{\v c}k{\' a}ch 2, 180 00 Praha 8, Czech Republic}

\author{Mokhtar Hassaine}
\email{hassaine@inst-mat.utalca.cl}
\affiliation{Instituto de Matem\'{a}ticas,
Universidad de Talca, Casilla 747, Talca, Chile}



\begin{abstract}

We construct explicit rotating solutions in Einstein's theory of relativity with a minimally coupled free scalar field rederiving and finding solutions in four or five spacetime dimensions. These spacetimes describe, in particular, the back-reaction of a free scalar field evolving in a Kerr spacetime. Adapting the general integrability result obtained many years ago from Eri\c{s}-G\"{u}rses to simpler spherical coordinates, we present a method for rederiving the four-dimensional Bogush-Gal'tsov solution. Furthermore, we find the five-dimensional spacetime featuring a free scalar with two distinct angular momenta. In the static limit, these five-dimensional geometries provide higher-dimensional extensions of the Zipoy-Voorhees spacetime. Last but not least, we obtain the four-dimensional version of a Kerr-Newman-NUT spacetime endowed with a free scalar, where the scalar field's radial profile is extended to incorporate dependence on the polar angular coordinate. Our results offer a comprehensive analysis of several recently proposed four-dimensional static solutions with scalar multipolar hair, representing a unified study of spacetimes with a free scalar field in both four and five dimensions under the general integrability result of Eri\c{s}-G\"{u}rses.


\end{abstract}
\maketitle

\section{Introduction}

The pedagogical significance of the Schwarzschild solution \cite{Schwarzschild:1916uq} and the astrophysical prominence of the Kerr geometry \cite{Kerr:1963ud} underscore the importance of studying static/stationary and axially symmetric spacetimes. These geometries, characterized by the presence of a pair of commuting Killing vectors $\xi=\partial_t$ and $\chi=\partial_\varphi$, describe the gravitational field of rotating sources that remain stationary in vacuum general relativity. Focusing on the stationary region outside the would-be event horizon, these geometries are in all generality described by the circular \textit{Weyl-Lewis-Papapetrou} (WLP) line element
\begin{equation}
\begin{aligned}
    ds^2&=-e^{{2}U(\rho,z)}[dt-\omega(\rho,z) d\varphi]^2\\
    &\quad+e^{-{2}U(\rho,z)}\left[e^{{2}\gamma(\rho,z)}(d\rho^2+dz^2)+\rho^2 d\varphi^2\right], \label{WLP}
\end{aligned}
\end{equation}
with coordinates such that $-\infty<t<\infty$, $0\leq\rho<\infty$, $-\infty<z<\infty$, and $0\leq\varphi<2\pi$. The function $\omega$ serves as a measure of rotation around the symmetry axis $\rho=0$, thereby representing the degree of stationarity of spacetime. Although these coordinates resemble standard cylindrical coordinates, they do not necessarily exhibit such symmetry when analyzing a particular geometry. Thus, at this level, $\rho$ and $z$ are treated as two spatial coordinates that simply yield a stationary and axially symmetric spacetime in the WLP form \eqref{WLP}. It is important to note that this line element can be directly extended to a configuration in electrovacuum by incorporating a gauge field that respects the corresponding symmetries dictated by $\xi$ and $\chi$, namely, $A=A_t(\rho,z)dt+A_\varphi(\rho,z)d\varphi$.

The field equations for the WLP form \eqref{WLP}, although intrinsically nonlinear, are known to be integrable. In fact, the vacuum field equations can be rearranged into a suitable form, known as the Ernst form \cite{Ernst:1967wx,Ernst:1967by}. This transformation reduces the problem to solving a single nonlinear differential equation for the complex Ernst gravitational potential. Once this equation is solved, it guarantees the integrability of all metric potentials present in \eqref{WLP}. Several solution-generating techniques have been developed based on the Ernst method \cite{Stephani:2003tm}, particularly due to the discovery of a family of Lie point symmetries of the Einstein system \cite{Ehlers:1957zz,Ehlers:1959aug,harrison1968new,kinnersley1973generation,Kinnersley:1977pg}, which remain obscure when not utilizing the complex Ernst potential framework.\footnote{The Ernst scheme \cite{Ernst:1967by,Ernst:1967wx} offers an alternative treatment of the electrovacuum. Two complex potentials $(\mathcal{E},\Psi)$, defined in terms of the metric functions appearing in \eqref{WLP}, allows to write the field equations in a manner in which a $SU(2,1)$ isometry group reveals transparent \cite{Stephani:2003tm}.}

The static case, where the ansatz \eqref{WLP} with $\omega=0$ reduces to the so-called Weyl metric, is considerably simpler. In fact, in vacuum, the field equations become
\begin{equation}
\nabla^2_{\mathbb{E}^3}U:=U_{,\rho\rho}+\frac{1}{\rho}U_{,\rho}+U_{,zz}=0, \label{Laplace}
\end{equation}
which is nothing but the Laplace equation in a three-dimensional Euclidean space in cylindrical coordinates. Quite remarkably the remaining metric function $\gamma$ carries all the nonlinearity of the Einstein equations and can be straightforwardly obtained from the following first-order equations, 
\begin{equation}
\gamma_{,\rho}=\rho\left(U^2_{,\rho}-U^2_{,z}\right),\quad  \gamma_{,z}=2\rho U_{,\rho}U_{,z}. \label{quadratures}
\end{equation}
This means that all static and axially symmetric solutions in vacuum are mathematically known (in a Weyl coordinate system). Indeed, the Laplace equation \eqref{Laplace} (which is linear) can be formally solved, and then the function $\gamma$ defined by means of \eqref{quadratures} can always be integrated. This establishes an interesting correspondence between axially symmetric Newtonian and stationary relativistic solutions. When the function $U$ is regarded as the Newtonian potential, it suffices to solve the Laplace equation \eqref{Laplace} for a given Newtonian source, then solve for $\gamma$ using the quadratures \eqref{quadratures}, thus obtaining the relativistic counterpart in the form of \eqref{WLP}. 
The Schwarzschild black hole \cite{Schwarzschild:1916uq} is known to be represented in the Weyl representation by a thin rod of mass $M$ and size $2M$, while a rod of the same mass but with an arbitrary linear density represents the so-called Zipoy-Voorhees spacetime \cite{Zipoy:1966btu,Voorhees:1970ywo}.  
Although the interpretation of the relativistic line element does not always follow directly from the Newtonian analog, insights from this approach have proven useful in constructing physically relevant spacetimes. Indeed, the linearity of the Laplace equation allows for the superposition of Newtonian potentials. Although modified by the nonlinearity introduced by the quadrature equations, this procedure has led, for example, to the construction of a collinear set of Schwarzschild black holes \cite{Israel:1964mgs}. The inherent nonlinearity in the equations for the function $\gamma$ introduces singular structures, referred to as struts (or conical singularities), which ultimately account for overcoming gravitational attraction between the individual bodies and keeping them apart. This system \cite{BachWeyl,Letelier:1997dz} is known to be connected with other important geometries, such as, for example, the C-metric line element \cite{Kinnersley:1969zz,Hong:2003gx}.

The situation is slightly different when considering matter fields, such as in electro-vacuum. The electromagnetic field introduces a nonlinearity in the equation for $U$, even in the static case. Nevertheless, stationary solutions can be derived, with the Reissner--Nordström \cite{Reissner:1916cle} and Kerr-Newman \cite{Newman:1965my} black holes serving as classic examples. Additionally, more sophisticated geometries can be constructed \cite{Belinski:2001ph,Stephani:2003tm}, primarily due to the Ernst method \cite{Ernst:1967by,Ernst:1967wx} and some of the Lie point symmetries it reveals \cite{harrison1968new}.

Exploring solutions involving matter fields beyond the electromagnetic case naturally brings us to the inclusion of a free scalar field energy-momentum matter tensor. We will also refer to this case as the Einstein-Scalar theory. This case has an interesting interpretation as it can be seen as the self-gravitating extension of a test scalar evolving in a given background geometry. The archetype example is a test scalar evolving in a Kerr geometry via the spin-$0$ Teukolsky equation \cite{Teukolsky:1973ha}. What would be the self-gravity of such a scalar induced on the spacetime geometry?
Furthermore, and this is central to our approach here, to what extent is the free scalar field source integrable in the framework of stationary and axially symmetric spacetimes. Via general conformal-disformal transformations, this would inform us of more complex cases involving Horndeski theories beyond the realm of GR \cite{Horndeski:1974wa,Gleyzes:2014dya,Gleyzes:2014qga,Langlois:2015cwa,Langlois:2015skt,Crisostomi:2016czh,BenAchour:2016fzp, BenAchour:2024hbg}. 

It turns out that in a mathematical framework, Eri\c{s} and G\"{u}rses have already answered the question of integrability since the seventies in a mostly forgotten publication. Indeed they very neatly demonstrated that the field equations of Einstein's theory coupled to a massless, minimally coupled scalar field possess the same degree of integrability as in the vacuum case \cite{Eris:1976xj}. In other words, for any stationary and axially symmetric solution of the vacuum Einstein equations, a corresponding solution with a real scalar field can always be constructed. This result is even more significant in the static case, where all static and axially symmetric vacuum spacetimes are already well-known. To be more specific concerning the result of \cite{Eris:1976xj}, we consider the Einstein-Scalar theory given by,
\begin{equation}
S_{{{\rm{ES}}}}=\int d^4x\sqrt{-g}\left(\frac{R}{4}-\frac{1}{2}\partial_\mu\Phi\partial^\mu\Phi\right),
\end{equation}
and whose field equations read
\begin{eqnarray} R_{\mu\nu}=2\partial_\mu\Phi\partial_\nu\Phi,\quad \Box\Phi=0.
\label{eqminscafield}
\end{eqnarray}
In Ref.  \cite{Eris:1976xj}, the authors established that any vacuum solution in the WLP form \eqref{WLP}, characterized by its defining functions $(U,\omega,\gamma)$, can be extended to a corresponding solution of the Einstein-Scalar system $(U,\omega,\gamma_{\rm{new}}=\gamma+\gamma^{\Phi},\Phi)$, provided $\gamma^{\Phi}$ and $\Phi$ satisfy the following equations
\begin{equation}    \gamma^{\Phi}_{,\rho}=\rho\left(\Phi^2_{,\rho}-\Phi^2_{,z}\right),\quad  \gamma_{,z}^\Phi=2\rho \Phi_{,\rho}\Phi_{,z},\quad  \nabla^2_{\mathbb{E}^3}\Phi=0. \label{quadratures2}
\end{equation}
The assumptions of stationarity and axial symmetry{\footnote{ There is a caveat to the above statement: in the presence of a free scalar one also needs to assume that the stationary and axisymmetric metric is a {\it{circular spacetime}}. In other words, the WLP Anzatz is no longer the most general axisymmetric and stationary metric and other solutions may well exist. We will not deal with such cases here.}} reduce the entire scalar field effect to the modification of the conformal factor $\gamma\rightarrow\gamma_{\rm{new}}$ of the two-dimensional planar sector $(\rho,z)$. This new nonlinear function is integrated via the quadratures \eqref{quadratures2}, once a solution of the Laplace equation for the scalar field $\Phi$ is selected. This is made possible primarily by the decoupling of the harmonic function $\Phi$. To be more complete, one can add that this theorem can be easily extended to cases involving a Maxwell source, provided the gauge field respects the same symmetries as the metric.

It should be noted that in accordance with no-hair theorems \cite{Herdeiro:2015waa} free scalar fields exhibit singular behavior and as such the spacetime metrics we shall be describing here are inherently singular. Another way to interpret this generic result is that the self-gravity of a free scalar field will generically render the event horizon singular. This is clearly because the scalar itself becomes singular as we approach the would-be horizon. Singular metrics are quite common even in vacuum GR, for example, the stationary Manko-Novikov class of metrics \cite{VSManko_1992} with differing mass-multipole moments (than those of Kerr). Given the phenomenological interest in stationary spacetime solutions albeit the difficulty in obtaining them, stationary ad hoc metrics have been constructed; some of them with inherent pathologies (see \cite{Psaltis:2023hvl} and references within). They have been studied extensively in tests of gravity as alternatives to the prototype of the Kerr black hole (see for example chapter 8 of \cite{Carson:2020gwh}). Generically such metrics are studied in regions away from their pathologies. However, and despite important scientific efforts, there is no known way to form a naked singularity from stable, astrophysically realistic initial conditions. Apart from the celebrated Penrose Cosmic Censorship Conjecture, this is the mathematical distinction between singular spacetimes and the Kerr metric.

On the other hand, general conformal/disformal transformations may convert singular horizons into regular ones when changing frames. A typical example of this is provided by the conformal extension of the Fisher-Janis-Newman-Winicour (FJNW) solution—the static and spherically symmetric solution of the Einstein-Scalar system \cite{Fisher:1948yn,Janis:1968zz}, given by the Bocharova-Bronnikov-Melnikov-Bekenstein black hole (BBMB) \cite{Bocharova:1970skc}. Furthermore, there are several ways known to circumvent no-hair theorems for scalars but they all involve adding terms in the Einstein-Scalar theory. For example, they may involve including higher order scalar tensor interaction terms that provide a source term for the free scalar such as those in \cite{Sotiriou:2013qea}. Alternatively, one can also break the symmetry of spacetime at the level of the scalar field (but not the associated energy-momentum tensor) in GR  \cite{Herdeiro:2014goa} or in modified theories of gravity (see for example \cite{Babichev:2023psy} and references within). 

Recently, starting from the solution found in \cite{Cardoso:2024yrb} the system of Eri\c{s} and G\"{u}rses has been independently studied in the static case  ($\omega=0$) in Ref. \cite{Herdeiro:2024oxn}, and some solutions representing Schwarzschild black holes immersed in specific scalar multipolar gravitational backgrounds have been discussed. Using $U= U_{{{ \rm{Schw}}}}$, where the Newtonian potential that gives rise to the Schwarzschild geometry is expressed in cylindrical coordinates as 
\begin{equation}
  U_{{{\rm{Schw}}}}(\rho, z)=\frac{1}{2} \ln \left[\frac{r_{+}+r_{-}-2 M}{r_{+}+r_{-}+2 M}\right], \,\, r_{ \pm}^2 \equiv \rho^2+(z \pm M)^2\nonumber,
\end{equation}
together with the general multipolar expansion solution for the scalar field $\Phi$ as given by 
\begin{equation}
\begin{aligned}    \Phi&=\frac{a_0}{\sqrt{\rho^2+z^2}}+\sum_{\ell=1}^{\infty}\left(\frac{a_{\ell}}{(\rho^2+z^2)^{\frac{(\ell+1)}{2}}}+b_{\ell} (\rho^2+z^2)^{\frac{\ell}{2}}\right)\\ &\quad\times P_{\ell}\left[\frac{z}{\sqrt{\rho^2+z^2}}\right],
\end{aligned}
\end{equation}
where $P_{\ell}$ are Legendre polynomials, and $a_{\ell}$ and $b_{\ell}$ are multipolar coefficients, and making use of \eqref{quadratures2} the scalar contribution $\gamma^{\Phi}$ can be integrated in full generality, yielding 
\begin{equation}
\begin{aligned}
&\gamma^\Phi=\sum_{\ell, m=0}^{\infty}\left[\frac{ \mathcal{A}_{\ell m} a_{\ell} a_m}{(\rho^2+z^2)^{\frac{\ell+m+2}{2}}}\left(P_{\ell+1} P_{m+1}-P_{\ell} P_m\right)\right]\\
&+\sum_{\ell, m=1}^{\infty}\left[\mathcal{B}_{\ell m} b_{\ell} b_m (\rho^2+z^2)^{\frac{\ell+m}{2}}\left(P_{\ell} P_m-P_{\ell-1} P_{m-1}\right)\right], \label{generalgamma}
\end{aligned}
\end{equation}
where we have defined the coefficients $\mathcal{A}_{\ell m}=\frac{(\ell+1)(m+1)}{\ell+m+2}$ and $\mathcal{B}_{\ell m}=\frac{\ell m}{\ell+m}$.
The expression \eqref{generalgamma} represents the gravitational scalar multipolar background in which $U= U_{{{\rm{Schw}}}}$ is immersed. Recall that, as a matter of fact, any other static vacuum solution $U$ can also be used.

It is well known that finding rotating solutions poses significant challenges; nonetheless, some solution-generating techniques have been developed to tackle these complexities effectively \cite{Bogush:2020lkp,Galtsov:1995mb,Clement:1997tx,Chauvineau:2018zjy}. Often the use of a particular coordinate system, like a Weyl coordinate system, is particularly well-tailored mathematically in order to achieve generality of results as we showed above. Nevertheless, such a set of coordinates may not be well adapted to certain physical cases and in particular obtaining explicit solutions. Therefore it is often necessary to slightly tweak generic methods to obtain simpler solutions in alternative coordinates. This is the route we will follow here, adapting the method of \cite{Eris:1976xj} to encompass certain new rotating solutions. Our approach is a simple enough example highlighting the value of alternative methods, which, though potentially less powerful or general, can significantly reduce computational effort and permit writing down explicitly certain solutions. 
We will provide a clear framework for integrating stationary and axially symmetric configurations within Einstein's theory coupled to a massless scalar field, avoiding reliance on Weyl coordinates. To demonstrate this, we begin by integrating the rotating extension of the FJNW solution, as presented in \cite{Bogush:2020lkp}. We also provide its charged and NUT generalizations with a more general scalar field dependence. This lays the groundwork for a novel application of our method: the construction, for the first time, of a scalar Myers-Perry-like configuration in five dimensions with two different angular momenta. 

The paper is organized as follows. In Section \ref{sec2}, we briefly review the four-dimensional rotating configuration of a scalar field minimally coupled to Einstein gravity as first presented in Ref. \cite{Bogush:2020lkp}. For later convenience (particularly in the five-dimensional case), we rederive this solution starting from the Kerr vacuum solution and applying a ``conformal transformation'' that acts solely on the diagonal and non-Killing sector of the metric, with a conformal factor that preserves the isometries of the original Kerr metric. In Section \ref{sec3}, this protocol is extended to the five-dimensional case, enabling the derivation of a rotating solution with two distinct angular momenta for a massless scalar field. It is also important to emphasize that, for these rotating solutions, the scalar field can be purely radial or, more generally, exhibit a dependence that respects the isometries of the metric. The emergence of this solution is a clear indication of the extension of the Eri\c{s} and G\"{u}rses framework in higher dimensions. We conclude in Section \ref{secclosing} by outlining further avenues for exploring the generation of these spacetimes through the application of our method to more sophisticated higher-dimensional rotating solutions. Additionally, two appendices are included. The first Appendix \ref{appA} aims to generalize the solution from \cite{Bogush:2020lkp} to the Kerr-Newman-NUT case and to develop an alternative treatment of the Eri\c{s} and G\"{u}rses equations, which naturally accommodates general scalar field profiles of the form $\Phi=\Phi(r,\theta)$. In the other Appendix \ref{appB}, we make use of the  Buchdahl theorems \cite{Buchdahl:1959nk,Barrientos:2024uuq} to construct the ZV-FJNW solution, facilitating the analysis of the static limit of our rotating configurations.
\section{Four-dimensional rotating solutions with a free scalar}\label{sec2}

In this section, we focus on the case of a massless scalar field minimally coupled to Einstein's gravity in four dimensions, as described by the equations \eqref{eqminscafield}. We begin with a brief overview of the solution obtained by Bogush and Gal'tsov \cite{Bogush:2020lkp}, which relies on the so-called Clement transformation \cite{Clement:1997tx}. Subsequently, we rederive this solution starting from the Kerr vacuum solution expressed in Boyer-Lindquist coordinates, 
 modified solely on the two-dimensional $(r,\theta)$ metric sector. Note that the metric remains unchanged in the Killing directions $(t,\varphi).$ 
 
 This second approach not only provides an alternative derivation but also lays the groundwork for extending the derivation to five dimensions.
\vspace{-0.2cm}
\subsection{The Bogush-Gal'tsov approach}
Constructing rotating solutions is usually more than a challenging task. A preliminary approach to developing a rotating generalization of the FJNW solution was proposed in \cite{Bogush:2020lkp}. In this reference, by employing the Clement transformation \cite{Clement:1997tx}, originally introduced to derive the Kerr metric from the Schwarzschild solution, an asymptotically flat rotating spacetime featuring scalar hair was successfully constructed. In more detail, the construction proceeds as follows. The starting point is to consider the Einstein-Scalar theory supplemented with a Maxwell term which is essential in order to extend the technique of Ref. \cite{Clement:1997tx} to the Einstein-Scalar framework. Using a WLP line element of the form \eqref{WLP}, the Einstein-Scalar-Maxwell system is shown to reduce to a three-dimensional $\sigma-$model, and the symmetries of the target space metric can be utilized, among other purposes, to introduce rotation to a given spacetime. First, the system is expressed in terms of the Ernst potentials $(\mathcal{E},\Psi)$. These potentials can be transformed into the Kinnersley form \cite{kinnersley1973generation}, where three new potentials $(U,V,W)$ are defined. The $SU(2,1)$ isometry group acts on these potentials, leaving invariant the norm $\bar{U}U+\bar{V}V-\bar{W}W=0$. The Clement transformation is defined using an appropriate combination of the target space discrete map $U \leftrightarrow V$ and a coordinate transformation that shifts the seed to a uniformly rotating frame. When applied to a type of FJNW configuration, and after some lengthy calculations, the following rotating spacetime configuration is obtained
\begin{widetext}
\begin{subequations}
\label{kerrmod}
\begin{align}
ds^2_{{{\rm{Rot.ZV-FJNW}}}} &= -\frac{\Delta(r)}{\varrho^2(r,\theta)}\left(dt-a\sin^2\theta d\varphi\right)^2 +
\frac{\sin^2\theta}{\varrho^2(r,\theta)}\left(a dt-(r^2+a^2)d\varphi\right)^2 + \frac{\varrho^2(r,\theta)}{\Delta(r)}H(r,\theta)\left(dr^2+\Delta(r)d\theta^2\right),\label{kerrmod1}\\
\Phi(r) &= \frac{\Sigma}{2\sqrt{M^2-a^2}}\ln\left(\frac{r-M-\sqrt{M^2-a^2}}{r-M+\sqrt{M^2-a^2}}\right)\label{kerrmod2},
\end{align}
\end{subequations}
\end{widetext}
where for simplicity, we have defined 
\begin{equation}
\Delta(r) = r^2 -2Mr+ a^2,\quad \varrho^2(r,\theta) = r^2 + a^2\cos^2\theta,
\label{def0}
\end{equation}
and 
\begin{equation}
\label{Hfunc}
H(r,\theta) = \left(1 + \frac{(M^2-a^2)}{\Delta(r)}\sin^2\theta\right)^{-\frac{\Sigma^2}{(M^2-a^2)}}.
\end{equation}
Here, $\Sigma$ represents the scalar charge associated with the scalar field, and $a$ stands for the rotation parameter.  We have expressed the solution presented in \cite{Bogush:2020lkp} in Boyer-Lindquist coordinates, clearly illustrating the backreaction of the scalar field on the line element, specifically by means of the metric function $H(r,\theta)$. This particular anzatz will be employed to produce an alternative derivation of the solution using the  Eri\c{s} and G\"{u}rses result. 

Before doing so it is important to note that for the correct application of the Clement transformation \cite{Clement:1997tx} in Einstein-Scalar theory, the presence of a Zipoy-Voorhees deformation parameter is necessary; otherwise, the residual electromagnetic field coming from the application of the  Clement transformation can not be removed. Consequently, the initial seed metric is not purely FJNW spacetime but rather its Zipoy-Voorhees extension (ZV-FJNW).\footnote{The ZV-FJNW solution is also presented in \cite{Bogush:2020lkp}. Recently, this metric has been integrated by brute force in \cite{Azizallahi:2023rrv} and generalized to include rotation \cite{Mirza:2023mnm}. As noted in \cite{Barrientos:2024uuq}, a straightforward construction of the ZV-FJNW can be achieved using Buchdahl's theorems \cite{Buchdahl:1956zz,Buchdahl:1959nk}.} Here, the deformation parameter is fixed in terms of the mass and scalar charge. Additionally, in the Einstein-Scalar framework, a Maxwell field has been incorporated; however, this field is eliminated in the final step of the construction through the symmetries of the Ernst potentials.

\subsection{Rederivation of the Bogush-Gal'tsov solution from the Kerr metric}
Let us now rederive the previous solution \eqref{kerrmod} via a simpler route using the aforementioned ansatz which boils down to the $H$ modification of the Kerr metric.
As mentioned earlier, according to the Eri\c{s} and G\"{u}rses result \cite{Eris:1976xj}, for any stationary and axially symmetric configuration, the complete effect of introducing the scalar field is encapsulated by the modification of the metric function $\gamma$. However, the direct application of this theorem necessitates the vacuum seed metric to be expressed in cylindrical coordinates, which can be quite cumbersome. With this in mind, it is easier to identify the backreaction of the scalar field on a given seed, enabling us to use more intuitive coordinates that facilitate the integration of the system.

In concrete, let us detail our method in order to easily derive the solution \eqref{kerrmod}. As a first step, we consider the vacuum Kerr line element expressed in Boyer-Lindquist coordinates. From \cite{Eris:1976xj}, we understand that the backreaction of the scalar field will modify the metric only in the non-Killing $(r,\theta)-$sector of the line element. We materialize this backreaction by a conformal factor $H(r,\theta)$ that acts on the relevant $(r,\theta)-$sector as expressed in \eqref{kerrmod1}. This metric ansatz \eqref{kerrmod1} has the nice property that the resulting metric remains Ricci flat along the Killing vector directions for any function $H(r,\theta)$, that is $R_{tt}=0$, $R_{t\varphi}=0$, and $R_{\varphi\varphi}=0$. In the second step, the remaining field equations will be determined by the dependence of the scalar field profile. In fact, for such a metric ansatz with a purely radial scalar field $\Phi=\Phi(r)$ the equations to be satisfied are solely $R_{r\theta}=0$, $R_{\theta\theta}=0$, and $R_{rr}=2\Phi_{,r}\Phi_{,r}$. Notably, from the off-diagonal equation $R_{r\theta}=0$, one yields 
\begin{equation} H(r,\theta)=F\left(\frac{\sin\theta}{\sqrt{r^2+a^2-2Mr}}\right),
\end{equation}
where $F$ denotes an arbitrary function of its argument. Injecting this expression into the equation $R_{\theta\theta}=0$, the form of $F$ is fully determined, providing us with 
\begin{equation}
H(r,\theta)=\left(1+\frac{(M^2-a^2)}{\Delta(r)}\sin^2\theta\right)^{-\frac{\Sigma^2}{(M^2-a^2)}},
\end{equation}
where $\Sigma^2$ is a positive integration constant. This is due to the remaining field equation $R_{rr}=2\Phi_{,r}\Phi_{,r}$ which restricts its sign according to 
\begin{equation}
 (\Phi^{\prime})^2\left(r^2+a^2-2Mr\right)^2-\Sigma^2=0.
\end{equation}
\\
The scalar field profile is therefore given by Eq. \eqref{kerrmod2}. In summary, in a few steps, we have easily obtained the solution \eqref{kerrmod}  without making use of cylindrical coordinates and their inherent complexities. A related derivation can also be applied to the case of an ansatz with a more general scalar field that depends on non-Killing coordinates. Indeed,  as we will show in Appendix \ref{appA}, solutions with a scalar field profile of the form $\Phi=\Phi(r,\theta)$ are perfectly attainable, despite the computational cost associated with solving the Klein-Gordon equation in a suitable gauge of the WLP metric in spherical coordinates. To facilitate this, we provide the Eri\c{s}-G\"{u}rses equations in spherical-like coordinates in Appendix \ref{appA}, where in this case, the flat Laplace equation is now represented by the standard curved Klein-Gordon equation. 

Some of the main features of the configuration \eqref{kerrmod} have already been discussed in \cite{Bogush:2020lkp}. However, for completeness, we highlight some of the most relevant aspects here and complement our analysis with some interesting limits. Firstly, one can note that the geometry described by the metric \eqref{kerrmod} is asymptotically flat, possesses a naked singularity at the would-be horizon $\Delta(r)=0$, and has the Kerr ring singularity located at $\varrho^2(r,\theta)=0$. Note that the position of the singular horizon remains unchanged. It is in fact unaffected by the scalar charge $\Sigma$, unlike the case of the Kerr-Newmann solution (for example). It is rather the scalar field that explodes at the horizon surface making spacetime singular in the process. The singular behavior can also be seen from the expression of the Ricci scalar that reads
\begin{equation}
    R=\frac{2\Sigma^2}{\Delta(r)\varrho^2(r,\theta)}\left(1+\frac{M^2-a^2}{\Delta(r)}\sin^2\theta\right)^{\frac{\Sigma^2}{M^2-a^2}}.
\end{equation}
All characteristics of the metric that have to do with the $(t,\varphi)$ sector, such as the ergosphere, presence of closed time curves, are identical to that of Kerr spacetime. 
In the case of zero scalar charge, $\Sigma=0$, we neatly obtain the Kerr solution while its static limit does not yield the Schwarzschild solution. In fact, the static limit $a=0$ reproduces a ZV-FJNW spacetime with a fixed value of the ZV-deformation parameter, specifically the one that relates the deformation parameter $\delta$ to the scalar charge $\beta$ such that $\delta\beta=1$ (see Appendix \ref{appB} for conventions and details). The static limit configuration yields
\begin{widetext}
\begin{subequations}
\label{statickerrmod}
\begin{align}
ds^2_{{{\rm{ZV-FJNW}}}}&=-\left(1-\frac{2M}{r}\right)dt^2+\left(\frac{1-\frac{2M}{r}+\frac{M^2\sin^2\theta}{r^2}}{1-\frac{2M}{r}}\right)^{-\frac{\Sigma^2}{M^2}}\left(\frac{dr^2}{\left(1-\frac{2M}{r}\right)}+r^2d\theta^2\right)+r^2\sin^2\theta d\varphi^2,\\
\Phi(r)&=\frac{\Sigma}{2M}\ln\left(1-\frac{2M}{r}\right).
\end{align}
\end{subequations}
\end{widetext}
The actual deformation is given by an overall parameter with a power $1-\delta^2=-\frac{\Sigma^2}{M^2}$. This is a natural consequence of the Eri\c{s} and G\"{u}rses theorem, as the scalar field modification of $\gamma$ invariably introduces a modification of a multipolar nature. It is worth noting that the procedure applied in \cite{Bogush:2020lkp} required a ZV extension of the FJNW to be implemented. Finally, the massless limit $M=0$ can be taken with care, and yields 
\begin{widetext}
\begin{equation}
ds^2 = -dt^2 + (r^2 + a^2) \sin^2 \theta \, d\varphi^2 + \left(\frac{r^2 + a^2 \cos^2 \theta}{r^2 + a^2}\right)^{1 + \frac{\Sigma^2}{a^2}} \left[dr^2 + (r^2 + a^2) d\theta^2\right], \qquad \Phi(r) = \frac{\Sigma}{a} \arctan\left(\frac{r}{a}\right).
\end{equation}
\end{widetext}
Notice that this metric is not flat mainly because of the backreaction of the scalar field.

\subsection{Charged rotating solution for a stationary and axisymmetric scalar field}
{
Now we consider the charged case with the action
\begin{equation}
S_{{{\rm{ESM}}}}=\int d^4x\sqrt{-g}\left(\frac{R}{4}-\frac{1}{2}\partial_\mu\Phi\partial^\mu\Phi-\frac{1}{4}F_{\mu\nu}F^{\mu\nu}\right),
\end{equation}
where as usual $F_{\mu\nu}=\partial_\mu A_\nu-\partial_\nu A_\mu$. The corresponding field equations are given by

\begin{equation}\label{einsteinscalarmaxwell}
\begin{aligned}
G_{\mu\nu}&=2\left(T^{\Phi}_{\mu\nu}+T^{F}_{\mu\nu}\right),\quad \square\Phi=0,\quad \nabla_\mu F^{\mu\nu}=0,
\end{aligned}
\end{equation}
where we have defined
{
\begin{equation}
\begin{aligned}
T^{\Phi}_{\mu\nu}&=\partial_\mu\Phi\partial_\nu\Phi-\frac{1}{2}g_{\mu\nu}\partial_\lambda\Phi\partial^\lambda\Phi,\\
T^{F}_{\mu\nu}&=F_{\lambda\mu}F^{\lambda}_{\,\,\,\,\nu}-\frac{1}{4}g_{\mu\nu}F_{\lambda\rho}F^{\lambda\rho}.
\end{aligned}
\end{equation}}}
Using the method that we outlined above it is now a simple exercise to include a $\theta$ dependence for the scalar field and to thus generalize the solution for a free scalar field which is axisymmetric and stationary. The new (to our knowledge) solution including electric charge reads, 
\begin{widetext}
    \begin{equation}
\begin{aligned}
ds^2&=-\frac{\Delta(r)}{\varrho^2(r,\theta)}\left(dt-a\sin^2\theta d\varphi\right)^2+
\frac{\sin^2\theta}{\varrho^2(r,\theta)}\left(a dt-(r^2+a^2)d\varphi\right)^2+\varrho^2(r,\theta)H(r,\theta)\left(\frac{dr^2}{\Delta(r)}+d\theta^2\right),\\
\Phi(r,\theta)&=\frac{\Sigma}{2\sqrt{M^2-a^2-e^2}}\ln\left(\frac{r-M-\sqrt{M^2-a^2-e^2}}{r-M+\sqrt{M^2-a^2-e^2 }}\right)+\frac{\Theta}{2\sqrt{M^2-a^2-e^2}}\ln\left(\frac{1-\cos\theta}{1+\cos\theta}\right),\\
A&=-\frac{er}{\varrho^2(r,\theta)}dt+\frac{aer\sin^2\theta}{\varrho^2(r,\theta)}d\varphi,
\label{kerrnewman}
\end{aligned}
\end{equation}
\end{widetext}
with
\begin{equation}
  \Delta(r)=r^2-2Mr+a^2+e^2,\,\,\,\, \varrho^2(r,\theta)=r^2+a^2\cos^2\theta. 
\end{equation}
The scalar solution is separable in $(r,\theta)$ and now has two independent charges $\Sigma$ and $\Theta$. 
The scalar field backreaction on the Kerr-Newmann metric is given exclusively by the $H$ function which now reads, 
\begin{widetext}
\begin{equation}
\begin{aligned}
H(r,\theta)&=\left(1+\frac{M^2-a^2-e^2}{\Delta(r)}\sin^2\theta\right)^{-\frac{\Sigma^2}{M^2-a^2-e^2}}\left(M^2-a^2-e^2+\frac{\Delta(r)}{\sin^2\theta}\right)^{-\frac{\Theta^2}{M^2-a^2-e^2}}\\
&\quad\times\left(\frac{r-M-\sqrt{M^2-a^2-e^2}\cos\theta}{r-M+\sqrt{M^2-a^2-e^2}\cos\theta}\right)^{\frac{2\Sigma\Theta}{M^2-a^2-e^2}}.
\end{aligned}
\end{equation}
\end{widetext}
In the appendix, we have given the above solution including the NUT charge $l$.

\section{Five-dimensional rotating solutions with two angular momenta and a free scalar field}\label{sec3}

Let us now address the construction of a five-dimensional rotating spacetime with a minimally coupled scalar field. While 
dimensional extensions of the Weyl problem have been studied \cite{Emparan:2001wk,Harmark:2004rm,Harmark:2005vn,Charmousis:2003wm}, there is little known regarding Einstein-Scalar theory. It may be tempting to directly apply the theorem of  Eri\c{s} and G\"{u}rses and evaluate its validity in higher dimensions. Explicitly, this would require working from the Myers-Perry black hole solution \cite{Myers:1986un}, expressed in Weyl canonical coordinates, which significantly complicates calculations. On the other hand, we have seen that in four dimensions, a slight modification of the vacuum metric in Boyer-Lindquist coordinates allows for a simple integration of the problem. This is precisely the approach we take here, and hence we consider the following metric ansatz
\begin{widetext}
\begin{equation}
\begin{aligned}
ds^2&=-\left(1-\frac{2Mr^2}{\varrho^2(r,\theta)}\right)dt^2+\frac{4 M r^2 (a \sin^2\theta d\xi+b\cos^2\theta d\chi)}{\varrho^2(r,\theta)} dt+\sin^2\theta\left(r^2+a^2+\frac{2Mr^2a^2\sin^2\theta}{\varrho^2(r,\theta)}\right)d\xi^2\\
&\quad+\cos^2\theta\left(r^2+b^2+\frac{2Mr^2b^2\cos^2\theta}{\varrho^2(r,\theta)}\right)d\chi^2+\frac{4Mr^2 ab \sin^2\theta\cos^2\theta}{\varrho^2(r,\theta)}d\xi d\chi+\frac{H(r,\theta) \varrho^2(r,\theta)}{\Delta(r)}\left[dr^2+\frac{\Delta(r)}{r^2}d\theta^2\right],
\label{sol5da2}
\end{aligned}
\end{equation}
\end{widetext}
with coordinates $(t,r,\theta,\xi,\chi)$, and with metric functions defined by 
\begin{equation}
\begin{aligned}
\varrho^2(r,\theta)&=r^2\left(r^2+a^2\cos^2\theta+b^2\sin^2\theta\right),\\
\Delta(r)&=(r^2+a^2)(r^2+b^2)-2Mr^2.
\end{aligned}
\end{equation}

In a complete analogy with the four-dimensional case,  for $H(r,\theta)=1$, this metric is nothing but the vacuum rotating Myers-Perry line element with two different angular momenta $a$ and $b$, \cite{Myers:1986un}. By following the same steps as in the four-dimensional case, we arrive at the conclusion that the metric function $H(r,\theta)$ is as follows
\begin{widetext}
\begin{equation}
H(r,\theta)=\left[\frac{4\left(r^2-M+\frac{a^2+b^2}{2}\right)^2-4\left(M-\frac{(a+b)^2}{2}\right)\left(M-\frac{(a-b)^2}{2}\right)
\cos^2(2\theta)}{r^4+(a^2+b^2-2M)r^2+a^2b^2}\right]^{\frac{-\alpha}{\left(2M-\vert a^2-b^2\vert\right)^2}},
\end{equation}
\end{widetext}
where $\alpha$ is an integration constant whose sign must be determined. On the other hand, the scalar field is given by the solution of the following nonlinear first-order equation
\begin{widetext}
\begin{eqnarray}
\left(2M-\vert a^2-b^2\vert \right)^2\left[r^4-r^2\left(2M-(a^2+b^2)\right)+a^2b^2\right]^2(\Phi^{\prime})^2=\alpha r^2\left[(a^2-b^2)^2+4M\left(M-(a^2+b^2)\right)\right].
\label{sfeqg}
\end{eqnarray}
\end{widetext}
It is clear and fully justified that both the function  $H(r,\theta)$ and the differential equation defining the scalar field must be invariant under the change $a\leftrightarrow b$. Assuming that (i) $M \in [0, \frac{(a-b)^2}{2}]$ or $M \geq \frac{(a+b)^2}{2}$ if $\text{sgn}(a) = \text{sgn}(b)$, or (ii) $M \in [0, \frac{(a+b)^2}{2}]$ or $M \geq \frac{(a-b)^2}{2}$ if $\text{sgn}(a) = -\text{sgn}(b)$, the scalar hair must be positive, $\alpha \geq 0$, allowing us to define $\alpha = \Sigma^2$. Thus, the scalar field reads
\begin{equation}
\Phi(r)=\frac{\Sigma}{2\left(2M-\vert a^2-b^2\vert\right)}\ln\left(\frac{2r^2-2M+a^2+b^2-\zeta}{2r^2-2M+a^2+b^2+\zeta}\right),
\label{sf2}
\end{equation}
where $\zeta\geq 0$ is given by
\begin{equation}
\zeta=\sqrt{(a^2-b^2)^2+4M\left[M-(a^2+b^2)\right]}. \label{deltasol}
\end{equation}
Finally, the metric function $H(r,\theta)$ can be compactly rewritten as 
\begin{widetext}
\begin{equation} H(r,\theta)=\left[4\left(1+\frac{\left(M-\frac{(a+b)^2}{2}\right)\left(M-\frac{(a-b)^2}{2}\right)}{\Delta(r)}\sin^2(2\theta)\right)\right]^{\frac{-\Sigma^2
}{\left(2M-\vert a^2-b^2\vert\right)^2}}.  
\end{equation}
\end{widetext}

In the previous discussion, it remains to consider the situation where the mass $M$ is bounded as
$$
M\in \, ]\frac{1}{2} (a\mp b)^2, \frac{1}{2} (a\pm b)^2[,
$$
with $\text{sgn}(a) =\pm \text{sgn}(b)$, which would correspond to the over spinning case. For this range of mass $M$, we have $P(M,a,b):=4M(M-(a^2+b^2))+(a^2-b^2)^2<0$ and, in this case, the integration constant $\alpha$ in \eqref{sfeqg} must be negative in order to deal with a real scalar field. The metric solution $ H(r,\theta)$ is then given by 
\begin{widetext}
\begin{equation} H(r,\theta)=\left[4\left(1+\frac{\left(M-\frac{(a+b)^2}{2}\right)\left(M-\frac{(a-b)^2}{2}\right)}{\Delta(r)}\sin^2(2\theta)\right)\right]^{\frac{\Sigma^2
}{\left(2M-\vert a^2-b^2\vert\right)^2}}, 
\end{equation}
\end{widetext}
while the scalar field reads
$$
\Phi(r)=\frac{\Sigma}{\left(2M-\vert a^2-b^2\vert\right)}\arctan\left(\frac{-2r^2-a^2-b^2+2M}{P(M,a,b)}\right).
$$
The scalar curvature is proportional to
$$
R\propto \Delta(r)^{\frac{\Sigma^2-(2M+\vert a^2-b^2\vert)^2}{(2M+\vert a^2-b^2\vert)^2}}.
$$
Although the scalar curvature does not diverge because $\Delta(r) > 0$ for this mass range, the Kretschmann invariant would exhibit singularities similar to those in the vacuum case, leading to a naked singularity.

The generalization of this solution with a more general scalar field $\Phi(r,\theta)$  can be straightforwardly achieved following Appendix \ref{appA}. The corresponding function $H(r,\theta)$ and scalar field profile are given by
\begin{widetext}
\begin{equation}
\label{generalscalarmp}
\begin{aligned}
H(r,\theta)&=\left(4\left[1+\frac{\left(M-\frac{(a+b)^2}{2}\right)\left(M-\frac{(a-b)^2}{2}\right)}{\Delta(r)}\sin^2(2\theta)\right]\right)^{-\frac{\Sigma^2
}{\left(2M-\vert a^2-b^2\vert\right)^2}}\\
&\quad \times \left(4\left[\frac{\Delta(r)}{\sin^2(2\theta)}+\left(M-\frac{(a+b)^2}{2}\right)\left(M-\frac{(a-b)^2}{2}\right)\right]\right)^{-\frac{\Theta^2
}{\left(2M-\vert a^2-b^2\vert\right)^2}}\\
&\quad\times \left(\frac{2r^2-2M+a^2+b^2-\zeta\cos(2\theta)}{2r^2-2M+a^2+b^2+\zeta\cos(2\theta)}\right)^{\frac{2\Sigma\Theta
}{\left(2M-\vert a^2-b^2\vert\right)^2}},\\
\\
\Phi(r,\theta)&=\frac{\Sigma}{2\left(2M-\vert a^2-b^2\vert\right)}\ln\left(\frac{2r^2-2M+a^2+b^2-\zeta}{2r^2-2M+a^2+b^2+\zeta}\right)+\frac{\Theta}{2\left(2M-\vert a^2-b^2\vert\right)}\ln\left(\frac{1-\cos(2\theta)}{1+\cos(2\theta)}\right),
\end{aligned}
\end{equation}
\end{widetext}
where $\zeta$ remains untouched with respect to the one given in \eqref{deltasol}. Note that this solution is characterized by two scalar charges given by $\Sigma$ and $\Theta$ while the scalar field solution is separable in sum with respect to the coordinates $r$ and $\theta$. 

Let us focus on the properties of the simple case with a purely radial scalar field, $\Phi=\Phi(r)$. The solution is asymptotically flat and exhibits curvature singularities at the ``horizon location'' $\Delta(r)=0$ and also at the ring $\varrho^2(r,\theta)=0$ as it can be seen from the Ricci scalar curvature invariant 
\begin{equation}
    R=\frac{32\zeta^2\Sigma^2r^2}{\left(2M-\vert a^2-b^2\vert\right)^2H(r,\theta)\Delta(r)\varrho^2(r,\theta)}.
\end{equation}
The zero scalar charge limit $\Sigma\rightarrow0$ reduces to the Myers-Perry configuration \cite{Myers:1986un}, while the static limit represents some interesting findings. Indeed, in the absence of angular momenta $a=b=0$, the configuration becomes
\begin{widetext}
\begin{subequations}
\label{staticmpmod}
    \begin{align}
        ds^2&=-\left(1-\frac{2M}{r^2}\right)dt^2+\left[\frac{4\left(1-\frac{2M}{r^2}+\frac{M^2\sin^2(2\theta)}{r^4}\right)}{1-\frac{2M}{r^2}}\right]^{\frac{-\Sigma^2}{4M^2}}\left(\frac{dr^2}{1-\frac{2M}{r^2}}+r^2 d\theta^2\right)+r^2\sin^2\theta d\xi^2+r^2\cos^2\theta d\chi^2,\\
        \Phi(r)&=\frac{\Sigma}{4M}\ln\left(1-\frac{2M}{r^2}\right),
    \end{align}
    \end{subequations}
\end{widetext}
and, as in the four-dimensional case, the scalar field introduces a multipolar modification encoded in the angular dependence of the scalar field backreaction $H(r,\theta)$. It is, therefore, natural to interpret this static limit as a five-dimensional ZV-FJNW spacetime \cite{Azizallahi:2023rrv}, where the deformation and scalar hair parameters are interconnected. It is worth noting that the lapse function takes the form of the five-dimensional Schwarzschild-Tangherlini spacetime \cite{Tangherlini:1963bw}, as expected, and that the scalar field is given by the logarithm of this function, as a naive application of Buchdahl transformations would suggest, in order to endow a five-dimensional spacetime with scalar hair. However, to the authors' knowledge, no higher-dimensional ZV spacetime in vacuum has yet been constructed in the literature, making it impossible to retrieve a general ZV-FJNW spacetime in arbitrary dimensions. 

\section{Closing remarks}\label{secclosing}

We have constructed new stationary and axially symmetric solutions for a scalar field minimally coupled to Einstein gravity in four and five dimensions. This system has been revisited in recent years, particularly in the static case, highlighting the ongoing interest in this framework. Indeed, in  Ref. \cite{Cardoso:2024yrb}, the authors have immersed a Schwarzschild black hole in a single monopole scalar environment, while in Ref. \cite{Herdeiro:2024oxn}, they have considered an immersion of a Schwarzschild black hole in the complete multipolar expansion associated with the solution of the Laplace equation for a static and axially symmetric scalar field configuration. On the other hand, in Refs. \cite{Mirza:2023mnm,Derekeh:2024hgh,Mazharimousavi:2023lxs}, several authors have explored the construction of ZV geometries with radially dependent scalar fields and with scalar field profiles depending exclusively on the polar angular coordinate, here denoted by $\theta$.

In the present paper, we relate all these solutions to the general result of Eri\c{s} and G\"{u}rses \cite{Eris:1976xj}. The multipolar modification of the resulting line elements is simply the effect produced by the scalar field, as can be directly obtained from \cite{Eris:1976xj}. In particular cases where the scalar field is purely radial, due to the form of the scalar field function $\Phi(r)\sim\ln\left(1-\frac{2M}{r}\right)$, it is clear that the modification induced on the metric function $\gamma$ of the ansatz metric \eqref{WLP} will take the ZV form, as this metric in vacuum is derived using a Newtonian potential exhibiting precisely the same form as $\Phi(r)$.

Furthermore, taking inspiration from the existing four-dimensional rotating solution in Einstein-Scalar theory \cite{Bogush:2020lkp} and its potential connection to the Eri\c{s} and G\"{u}rses \cite{Eris:1976xj}, we extrapolated the five-dimensional analog, accounting for the additional complexity of two angular momenta. In doing so, we introduced an alternative method addressing the Eri\c{s} and G\"{u}rses result while effectively bypassing the use of Weyl coordinates. Indeed Weyl coordinates in certain cases incur a high computational cost when dealing with rotating metrics. We thus found, for the first time, a five-dimensional Einstein-Scalar metric with two different angular momenta. Due to the intrinsic connection between the Eri\c{s} and G\"{u}rses mechanisms and the embedding of geometries in multipolar scalar environments, we observe that in our five-dimensional case, the static limit provides a ZV-FJNW line element in five dimensions where the ratio between the deformation parameter and the scalar charge is fixed.

In addition, our method permitted us to generalize these findings to the case where the scalar field profile takes a full stationary, axisymmetric form $\Phi(r,\theta)$.\footnote{We note however, that we have not managed to show that our separable form of the scalar is not the most general stationary and axisymmetric scalar solution.} This framework has also been applied to the four-dimensional solution \cite{Bogush:2020lkp} to derive its Kerr-Newman-NUT generalization featuring general scalar hair. It similarly extends to all solutions discussed in \cite{Mirza:2023mnm,Derekeh:2024hgh,Mazharimousavi:2023lxs}, enabling the construction of additional static, higher-dimensional, axially symmetric solutions with multipolar scalar modifications. Notably, this approach avoids the need to prioritize either the radial or polar dependence of the scalar field profile.

Based on our findings, it is clear that there are several avenues to explore that could prove to be interesting. 
The first one is the study of conformal and disformal configurations of our Einstein-Scalar system which will lead to some form of Horndeski theory \cite{BenAchour:2016cay}. Important steps have been taken for the case of a conformal coupling where rotating versions of the BBMB black hole have been constructed using solution-generating methods \cite{Astorino:2014mda}. When the scalar is conformally coupled one expects better spacetime regularity, as in this frame the spacetime horizon is not singular. The scalar is still expected to be singular on the locus of the spacetime horizon.
Regularity for a conformally coupled scalar is improved in the presence of a cosmological constant as the additional length scale permits to differentiate the horizon locus \cite{Martinez:2002ru}. However, it is well-known that the cosmological constant spoils the integrability properties of the WLP equations \cite{Charmousis:2006fx} (see the extension of the Ernst method in \cite{Astorino:2012zm}). In particular, it is straightforward to see that the Eri\c{s}-G\"{u}rses result does not encompass the case of a cosmological constant. Equally evident is that applying a conformal transformation in the non-Killing sector of the Kerr-(A)dS solution will not work in this context. We believe that, in addition to introducing a cosmological constant into the setup, one should also consider a self-interaction potential and investigate whether it would be possible to construct rotating solutions.

The extension to the case of a complex scalar field proves equally challenging, partly because the equations that highlight the modulus and phase of the scalar field do not decouple as they do in the real case. Numerically, this problem has been considered in four dimensions with great success \cite{Herdeiro:2014goa}, but explicit solutions have not been found. Nevertheless, as with the case of the cosmological constant, this remains a problem worth investigating. The question of finding a charged solution in five dimensions with a scalar field is equally interesting. It is known that in the absence of the scalar field, it is possible to charge the solutions through Maxwell term, but this must be supplemented by the Chern-Simons term to obtain rotating charged solutions \cite{Chong:2005hr}, which reduce to the Myers-Perry solution in the absence of charge \cite{Myers:1986un}. Also, when thinking of five dimensions, one immediately thinks of black rings \cite{Emparan:2001wn} and hence the possibility for a scalar field to be described by an extension of the black ring metric. Recently, numerical solutions describing static, asymptotically flat black rings were presented in \cite{Kleihaus:2019wck}, but for a phantom scalar field, meaning that the kinetic term of the scalar field has the opposite sign. On the other hand, since the vacuum black ring solution is stationary and axisymmetric, it should be possible to apply the generalization of the Eri\c{s}-G\"{u}rses theorem. However, it remains unclear whether analytical expressions for the solution can be obtained. This is due in part to the fact that the black ring solution is conveniently written through prolate coordinates, and it is not immediately clear how to formulate our problem using these coordinates. {As it is well-known, solutions of a scalar field minimally coupled to Einstein gravity can be brought to the Brans-Dicke frame by performing a conformal transformation using the scalar field. Nevertheless, a no-hair theorem applies since our models do not include self-interactions for the scalar field, $V=0$, neither in the Einstein frame nor in the Jordan frame. In particular, it is known that in Brans-Dicke theory without a scalar potential, stationary black hole solutions must have a constant scalar field and reduce to those of general relativity \cite{Hawking:1972qk}. It is clear from the  Eri\c{s} and G\"{u}rses 
 theorem that starting from any solution of the Klein-Gordon equation, one can generate a corresponding solution of the Einstein-scalar field theory, leading to a large set of possibilities. In consequence, an equivalently large spectrum of solutions exists in all those models related to the Einstein-scalar model, as for example, the case of Brans-Dicke theory. Some of these solutions have already been presented in the literature, see for example \cite{Singh:1979va,Kim:1998hc,Kim:2004pxa,Kirezli:2015wda}.}



\acknowledgments

The work of J.B. is supported by FONDECYT Postdoctorado grant 3230596. A.C. is partially supported by FONDECYT grant 1210500 and by GA{\v C}R 22-14791S grant of the Czech Science Foundation.  The work of M.H. is partially supported by FONDECYT grant 1210889. We would like to thank Nicolas Lecoeur for the discussions we had with him at the early stage  of this work. C.C. thanks Karim Noui and Georgios Pappas for enlightening discussions.

\appendix
\section{Field equations in spherical-like coordinates and charged extension}\label{appA}
As previously emphasized, solutions with a general scalar field profile of the form $\Phi=\Phi(r,\theta)$ are achievable. However, both our protocol and the treatment by Eri\c{s} and G\"{u}rses involve significant computational costs. The latter proves to be cumbersome, particularly due to the employing of cylindrical coordinates. Nevertheless, the field equations of stationary and axially symmetric configurations in Einstein-Scalar theory remain manageable when an appropriate gauge for the WLP configuration is chosen. This is what we propose to present in the next sub-section.

\subsection{Eri\c{s} and G\"{u}rses like equations in spherical-like coordinates}
 In fact, using the following spherical-like ansatz for the metric (which is a mix of the WLP and spherical symmetry)
\begin{equation}\label{LWPspherical}
\begin{aligned}
ds^2&=-e^{2U}[dt-\omega d\varphi]^2\\ &\quad+e^{-2U}\left[e^{2\gamma}\left(\frac{dr^2}{\Delta_1}+\frac{ d\theta^2}{\Delta_2}\right)+\varrho^2 d\varphi^2\right],
\end{aligned}
\end{equation}
where the functions $U,\,\omega,\,\gamma,\,\varrho^2$ depend on $(r,\theta)$ and, where $\Delta_1=\Delta_1(r)$ and $\Delta_2=\Delta_2(\theta)$, the scalar field backreaction is described by a solution to the system
\begin{equation}\label{equationsgursesspherical}
\begin{aligned}
\left(\ln \varrho^2\right)_{,r}\gamma^{\Phi}_{,\theta}+\left(\ln \varrho^2\right)_{,\theta}\gamma^{\Phi}_{,r}&=4\Phi_{,r}\Phi_{,\theta},\\
\Delta_{1}\left(\ln \varrho^2\right)_{,r}\gamma^\Phi_{,r}-\Delta_2\left(\ln \varrho^2\right)_{,\theta}\gamma^{\Phi}_{,\theta}&=2(\Delta_1 \Phi_{,r}^2-\Delta_2\Phi_{,\theta}^2).
\end{aligned}
\end{equation}
Notice that, in these coordinates, the quadratures include the seed metric functions. Additionally, a solution for $\Phi$ is not fully assured, as in these coordinates, the scalar field is governed not by a Laplace equation in flat cylindrical coordinates but by the standard curved Klein-Gordon equation. However, for a line element of the form \eqref{LWPspherical}, the Klein-Gordon equation is solvable in several cases of interest. Lastly, we note that a similar method was used to obtain the Ernst potentials for certain solutions including a cosmological constant \cite{Charmousis:2006fx}.

\subsection{The Kerr-Newman-NUT configuration with scalar hair  and solutions with \texorpdfstring{$\Phi=\Phi(r,\theta)$}{Lg}}
A Kerr-Newman-NUT generalization of \eqref{kerrnewman} can be easily obtained, yielding
\begin{widetext}
    \begin{equation}
\begin{aligned}
ds^2&=-\frac{\Delta(r)}{\varrho^2(r,\theta)}\left(dt-(a\sin^2\theta+2l\cos\theta) d\varphi\right)^2+
\frac{\sin^2\theta}{\varrho^2(r,\theta)}\left(a dt-(r^2+a^2+l^2)d\varphi\right)^2+\varrho^2(r,\theta)H(r,\theta)\left(\frac{dr^2}{\Delta(r)}+d\theta^2\right),\\
\Phi(r,\theta)&=\frac{\Sigma}{2\sqrt{M^2+l^2-a^2-e^2}}\ln\left(\frac{r-M-\sqrt{M^2+l^2-a^2-e^2}}{r-M+\sqrt{M^2+l^2-a^2-e^2 }}\right)+\frac{\Theta}{2\sqrt{M^2+l^2-a^2-e^2}}\ln\left(\frac{1-\cos\theta}{1+\cos\theta}\right),\\
A&=-\frac{er}{\varrho^2(r,\theta)}dt+\frac{er(a\sin^2\theta+2l\cos\theta)}{\varrho^2(r,\theta)}d\varphi,
\label{kerrnewmannut}
\end{aligned}
\end{equation}
\end{widetext}
with
\begin{equation}
  \Delta(r)=r^2-2Mr+a^2-l^2+e^2,\, \varrho^2(r,\theta)=r^2+(l-a\cos\theta)^2,   
\end{equation}
and where the scalar field backreaction is given by 
\begin{widetext}
\begin{equation}
\begin{aligned}
H(r,\theta)&=\left(1+\frac{M^2+l^2-a^2-e^2}{\Delta(r)}\sin^2\theta\right)^{-\frac{\Sigma^2}{M^2+l^2-a^2-e^2}}\left(M^2+l^2-a^2-e^2+\frac{\Delta(r)}{\sin^2\theta}\right)^{-\frac{\Theta^2}{M^2+l^2-a^2-e^2}}\\
&\quad\times\left(\frac{r-M-\sqrt{M^2+l^2-a^2-e^2}\cos\theta}{r-M+\sqrt{M^2+l^2-a^2-e^2}\cos\theta}\right)^{\frac{2\Sigma\Theta}{M^2+l^2-a^2-e^2}}.
\end{aligned}
\end{equation}
\end{widetext}
Here, $\Theta$ is another integration constant that, along with $\Sigma$, represents the scalar hair. It is important to note that this strategy also applies in five dimensions, yielding the scalar field profile presented in \eqref{generalscalarmp}.

\section{The ZV-FJNW spacetime}\label{appB}

In spherical coordinates $-\infty<t<\infty$, $0\leq r<\infty$, $0<\theta<\pi$ and $0\leq\varphi<2\pi$, the Zipoy-Voorhees spacetime reads \cite{Esposito:1975qkv,Lukes-Gerakopoulos:2012qpc} 
\begin{equation}
    ds^2_{{{\rm{ZV}}}}=-f^\delta dt^2+\frac{\left[\left(\frac{f}{g}\right)^{\delta^2}g\left(\frac{dr^2}{f}+r^2d\theta^2\right)+fr^2\sin^2\theta d\varphi^2\right]}{f^\delta},
\end{equation}
with
\begin{equation}
    f=\left(1-\frac{2M}{r}\right),\quad g=\left(1-\frac{2M}{r}+\frac{M^2\sin^2\theta}{r^2}\right),
\end{equation}
and where $\delta$ represents the deformation parameter. Its Newtonian interpretation is that of a finite thin rod of mass $M$ and length $2\ell$, with a linear density $\sigma=\frac{\delta}{2}$. Since it is a static vacuum solution, it is subject to the application of Buchdahl's theorem of the second kind \cite{Buchdahl:1959nk,Barrientos:2024uuq} and, therefore, can be dressed with a minimally coupled scalar field. This configuration results in the ZV-FJNW spacetime
\begin{equation}
\begin{aligned}
     ds^2_{{{\rm{ZV}}}}&=-f^{\delta\beta} dt^2+\frac{\left[\left(\frac{f}{g}\right)^{\delta^2}g\left(\frac{dr^2}{f}+r^2d\theta^2\right)+fr^2\sin^2\theta d\varphi^2\right]}{f^{\delta\beta}},\\
     \Phi&=\frac{\delta}{2}\sqrt{1-\beta^2}\ln\left(1-\frac{2M}{r}\right),
\end{aligned}
\end{equation}
where $\beta$ is the scalar charge associated with the scalar field. In the particular case of $\delta\beta\rightarrow1$, the spacetime reproduces the static limit of \eqref{kerrmod} discussed in \eqref{statickerrmod}. There, we identify $1-\delta^2$ with $-\frac{\Sigma^2}{M^2}$.
\bibliography{apssamp}
\end{document}